# Mining Misdiagnosis Patterns from Biomedical Literature

Cindy Li[1], Elizabeth Chen, PhD[1], Guergana Savova, PhD[2], Hamish Fraser, MBChB[1], Carsten Eickhoff, PhD[1]
[1]Center for Biomedical Informatics, Brown University, Providence, RI, United States
[2]Computational Health Informatics Program, Boston Children's Hospital and Harvard Medical School, Boston, MA, United States

**Abstract**

*Diagnostic errors can pose a serious threat to patient safety, leading to serious harm and even death. Efforts are being made to develop interventions that allow physicians to reassess for errors and improve diagnostic accuracy. Our study presents an exploration of misdiagnosis patterns mined from PubMed abstracts. Article titles containing certain phrases indicating misdiagnosis were selected and frequencies of these misdiagnoses calculated. We present the resulting patterns in the form of a directed graph with frequency-weighted misdiagnosis edges connecting diagnosis vertices. We find that the most commonly misdiagnosed diseases were often misdiagnosed as many different diseases, with each misdiagnosis having a relatively low frequency, rather than as a single disease with greater probability. Additionally, while a misdiagnosis relationship may generally exist, the relationship was often found to be one-sided.*

**Introduction**

Diagnostic errors can be frequent, costly, and sometimes fatal in medicine. It is one of the most pressing issues on patients' minds, with 22% of all patients seeking emergency treatment expressing concern over misdiagnoses[1]. In a randomized survey, more than 1 in 10 respondents reported having experienced an issue with their diagnosis in the past. It is estimated that about 12 million Americans may face a diagnostic error each year, and half of these errors have the potential to cause serious harm[2]. Worse, according to one study, 83.3% of diagnostic errors were preventable[3].

Diagnostic errors can cause serious costs for both the healthcare provider and the patient. Medical errors, among which diagnostic errors were found to make up around 28.7% based on malpractice claims, cost the United States $19.5 billion in 2008 alone[4,5]. The extra cost for additional treatment, more deaths, and lost productivity is driving up the cost of healthcare, hurting both healthcare providers and patients seeking treatments[5]. Malpractice claims are also common and costly, with sometimes exorbitant litigation fees[6]. Diagnostic errors contribute a significant portion to these cases, ranging from 30 to 59%[7,8].

For patients, the consequences of diagnostic errors are even more direct. In 59% of the malpractice claims resulting from diagnostic errors, serious harm was caused by these errors and 30% resulted in death[7]. Even in less serious cases, diagnostic errors may lead to patients needing more doctor visits, longer hospital stays, or receiving inappropriate treatment[9,10].

Physicians can be affected by cognitive biases and personality traits when making medical decisions. Factors such as overconfidence or risk aversion can affect the accuracy of their diagnoses. These cognitive biases were associated with inaccuracies in diagnosis in 36.5% to 77% of case scenarios[11]. Furthermore, physicians are often unaware of mistakes they make. Interventions that force doctors to reassess and be more aware of possible errors, may thus be very useful in reducing the frequency of diagnostic errors[8].

One study described an intervention that has already shown to improve diagnostic accuracy through a computerised diagnostic support system, where physicians code in symptoms, and a list of possible diagnoses along with their likelihood is returned[12]. In another study, 74% of physicians felt that these systems were useful, helping them consider more diagnoses and ask more specific questions[13]. Furthermore, these systems did not increase the duration of the consultation nor the number of tests ordered. Yet another study's system showed an improvement of 8-9% in diagnostic accuracy, which, given the many consultations that occur every day, would benefit a significant number of patients[12].

We hypothesize that these approaches may be improved by integrating patterns of misdiagnosis, allowing them to return not only the likelihood of a diagnosis but also the likelihood of its misdiagnosis as another disease. Currently, some systems are already considering cases of misdiagnosis. One model plans to scan the records of the hospitals using the system to determine whether a misdiagnosis occurred and store the misdiagnosis as an attribute that would affect the weights of symptoms of the given disease when determining a diagnosis[14].

In this study we explore accounts of misdiagnoses in the PubMed database as a resource for potentially flagging misdiagnosis.

**Methodology**

We used the PubMed database, specifically all articles contained in the PubMed 2018 annual baseline (N=27,837,540 citations). Each article is an XML file with title, author, and abstract tags among others. Using Python, we parsed through these articles, selecting only those whose titles contained the phrases, "misdiagnosed as" or "masquerading as". These phrases were chosen after testing phrases using the PubMed database search tool to determine which ones produced the most results. Titles were examined for relevancy and those that did not pertain to medical terms were filtered out when matching.

From the selected titles, we used QuickUMLS[15], a medical named entity resolution library, to extract disease names and their Universal Medical Language System concept unique identifiers (UMLS CUIs), each of which is denoted by a "C" followed by 7 digits, selecting specifically for semantic types T047 (Disease or Syndrome) and T191 (Neoplastic Process). QuickUMLS returned a list of all matches, with each entry in the list representing a list of all possible matches to a medical term found in the title. Each match was represented by a dictionary with the start and end of the medical term being matched, the UMLS CUI, the term associated with the UMLS CUI, the degree of similarity from 0 to 1, and the UMLS semantic type. Since a given medical term may match with multiple CUIs, we prioritized the longest match, followed by the greatest similarity, followed by the CUI with the lowest 7-digit number. We considered only titles that contained exactly two matches, one preceding our selected phrase and one following after it. Figure 1 illustrates this process.

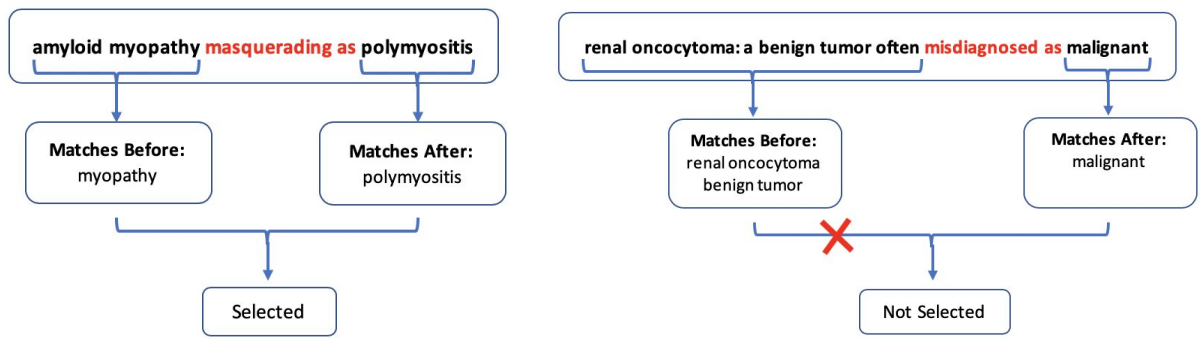

**Figure 1.** Examples of titles matching our condition (left) and not matching our condition (right). The selection criterion was that there is only a single medical concept recognized preceding the chosen phrase (red) and only a single one following after it. For the title on the right, there were two recognized medical concepts preceding the chosen phrase, so the title was not selected.

The observation frequencies of each disease pair and of each disease being misdiagnosed as another were calculated. Using the UMLS API[16], each CUI extracted was checked for parent/child or synonymy relationships with other CUIs. If a CUI had a parent-child (PAR or RB and CHD or RN, respectively) or synonymous (SYN or RL) relationship with another CUI extracted from one of our titles, it was counted as that parent or synonymous CUI[17] (Figure 2). Each CUI pair frequency was then normalized by the frequency of the first CUI, which represents the correct diagnosis, and a graph was generated with the normalized frequencies as the weights of the directed edges. The graph was created and graph statistics analyzed using NetworkX[18].

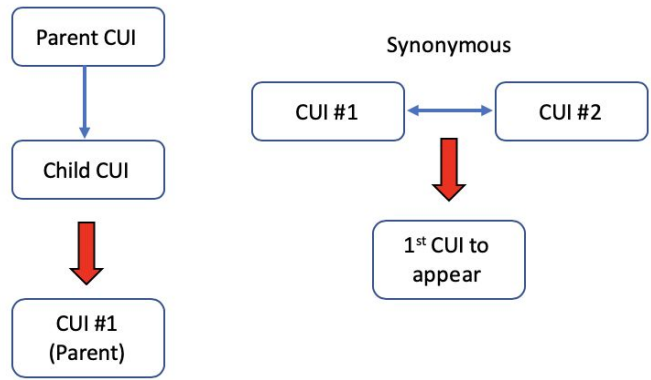

**Figure 2.** Diagram of parent/child and synonymy CUI relationships and the resulting chosen CUI. In the case of parent/child relationships, the parent CUI was always chosen. For synonymy relationships, the first synonymous CUI to appear in the selected titles was chosen.

**Results**

Of 5,105 titles that contained our given phrase, 2,502 misdiagnosis pairs were extracted. Each node represents a unique CUI and the arrows indicate a misdiagnosis relationship, where the source node is the correct diagnosis, and the destination node is the incorrect diagnosis. The darker node indicates that the source node was more often misdiagnosed as the destination node than any of its other destination nodes.

The larger nodes in the center indicate that several CUIs are misdiagnosed as other diseases relatively frequently whereas the smallest nodes along the edges indicate that there are diseases that were rarely or never reported misdiagnosed based on the articles extracted. The darker edges suggest that when the source node is misdiagnosed, it is more often misdiagnosed as the destination node; however, in many cases, especially in the case of the nodes on the edge of the graph, the greater weight is due to the frequency of the source node being very low to begin with (Figure 3).

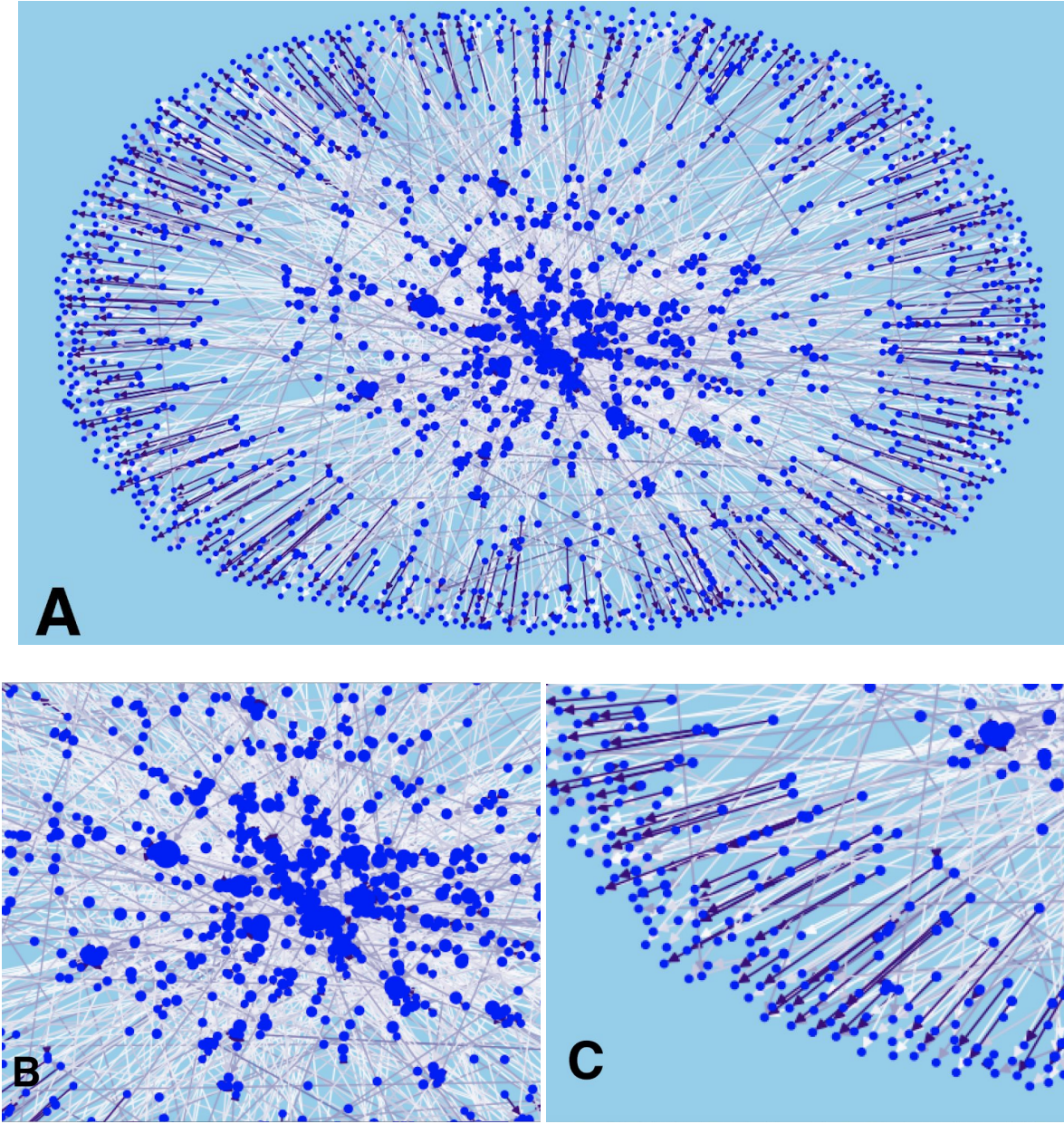

**Figure 3.** Graphs of all CUIs and misdiagnosis relationships extracted. Graph is drawn using a force-directed algorithm. The node closest to the arrowhead is the disease that another is being misdiagnosed as. A larger node means a greater out-degree centrality while a darker arrow means the relationship has a greater normalized frequency. (A) All nodes shown. (B) Nodes in the center tend to be larger. (C) Nodes along the edge tend to be smaller.

While many diseases only had a few incidences of misdiagnosis cited in the literature, several were misdiagnosed relatively frequently, the top five of which can be found in Table 1. The single most commonly misdiagnosed disease was tuberculosis (C0041296), which had both the greatest source frequency, or number of mentions of being misdiagnosed, as well as the greatest out-degree (i.e. the number of different diseases it was wrongly diagnosed as). This is not surprising from a clinical viewpoint, tuberculosis has long been known as a mimic of other diseases. It was most often misdiagnosed as a form of carcinoma followed by malignancies and other tumors. The frequency with which tuberculosis was misdiagnosed as each disease were not high, though, with tuberculosis being misdiagnosed as carcinoma in about 10% of cases of tuberculosis misdiagnosis.

Similarly, for the other most commonly misdiagnosed diseases, the frequency of each misdiagnosis pair was fairly low, with the highest being about 13% of all cases of that disease being misdiagnosed (Table 1). Overall, the normalized frequencies were relatively low, except in the cases where the occurrence of the correct diagnosis was low to begin with.

**Table 1**. Top 5 most commonly misdiagnosed diseases according to our results, along with their out-degree, source frequency, or how often a given disease was misdiagnosed as another, and the disease they are most mistaken as.

| Name (CUI) | Source Frequency | Out Degree | Most Frequently Mistaken as (CUI) & Frequency |
|---|---|---|---|
| Tuberculosis (C0041296) | 68 | 46 | Carcinoma (C0041296) 0.1029 |
| Cyst (C0010709) | 44 | 34 | Pericarditis (C0031046) 0.0455 |
| Tumor (C0027651) | 40 | 37 | Liver Secondaries (C0494165) 0.0500 |
| Nerve Sheath Tumor (C0206727) | 32 | 24 | Carotid Body Tumor (C0007279) 0.1250 |
| Neuroendocrine Tumor (C0206754) | 31 | 28 | Diabetic Foot Ulcer (C1456868) 0.1290 |

When inspecting reversed disease pair relationships, it was found that while one disease may have been most frequently misdiagnosed as another, the same did not often hold in the reverse direction. For example, while tuberculosis was most frequently misdiagnosed as a carcinoma, there were no mentions at all of carcinoma misdiagnosed as tuberculosis in our results. Of the top 20 most misdiagnosed diseases, for all but two diseases, the CUI that the disease was most misdiagnosed as was never misdiagnosed as the disease. In those two other cases, the frequency only went up to about 8.7%.

Just as with the misdiagnosed diseases, most of the diagnosed diseases that were later found to be incorrect only had a few incidences, although a few were relatively often diagnosed when it was that disease. The disease with highest in-degree, or the one that other diseases are most commonly misdiagnosed as, are tumors (C0027651). As with the misdiagnosed diseases, though, when a disease was incorrectly diagnosed, there was no single disease that was very frequently the correct diagnosis. In only about 8.21% of cases where a tumor was a destination node, or incorrect diagnosis, the correct diagnosis was tuberculosis and this was the highest frequency for tumors. For the other most often incorrectly diagnosed diseases, the frequency only reached about 12% of all cases where that disease was the incorrect diagnosis (Table 2).

**Table 2.** Top 5 diseases most often wrongly diagnosed according to our results, along with their in-degree, destination frequency, or how often other diseases were misdiagnosed as a given disease, and the most frequent correct diagnosis.

| Name (CUI) | Destination Frequency | In Degree | Most Frequent Correct (CUI) & Frequency |
|---|---|---|---|
| Tumor (C0027651) | 73 | 62 | Tuberculosis (C0041296) 0.0821 |
| Malignancy (C0006826) | 59 | 45 | Actinomycosis (C0001261) 0.1186 |
| Cyst (C0010709) | 42 | 31 | Connective Tissue Tumor (C0027656) 0.0952 |
| Tuberculosis (C0041296) | 35 | 33 | Carcinoma of Lung (C0684249) 0.0857 |
| Skin Conditions (C0037274) | 34 | 30 | Skin Ulcers (C0037299) 0.1176 |

As with the normalized frequencies, the frequency of each disease given the incorrect diagnosis was relatively low for most misdiagnoses except in cases where the wrong disease had a low frequency. To facilitate further research, we share all extracted misdiagnosis pairs alongside the computed normalized frequencies and graphs with the research community (https://github.com/bcbi-edu/p_eickhoff-misdiagnosis).

**Discussion**

Our study shows that diagnoses are often mistaken for a variety of diseases rather than a single recurring one. A disease pair often occurred only once, and for disease pairs with a frequency greater than one, the normalized count was often low. Thus, the frequency of the disease being misdiagnosed as a single specific disease was relatively low. Likewise, incorrect diagnoses link to multiple correct diagnoses.

This one-sided directionality of misdiagnosis relationships suggests that while symptoms for the two diseases may be similar, leading to the first disease often being misdiagnosed as the second, the second disease may not be frequently misdiagnosed as the first. This may in some cases be due to the rarity of the diseases.

In our current analysis, much data (about 51%) was lost due to the often complicated structures of the article titles. More data may also be gained by looking at more than just article titles when selecting those to evaluate initially as articles may discuss misdiagnoses without explicitly stating it in the title. In this regard, abstracts, full texts and annotated resources such as SemRep[19] or SemMedDB1[20] will be of great value.

Our work provides a first step towards obtaining misdiagnosis patterns and frequencies. Future work may include factoring in overall prevalence rates of diseases instead of just cases where the diagnosis was erroneous. Our observation frequencies currently only reflect the probability of one disease given an incorrect diagnosis. Healthcare providers may benefit more from also knowing the frequency of a diagnosis being incorrect. Additionally, studying orthogonal sources, possibly open medical records or malpractice claims, as well as additional phrases indicating misdiagnoses may offer more insights.

In the long run we are excited to use these early insights towards diagnostic decision support. For example, if meningitis is often misdiagnosed as the flu - then that should direct us to ensure that patients with an initial diagnosis of flu receive the correct screening questions to help discriminate them. One also might use this data to devise more effective and cost efficient strategies for investigation of patients with certain provisional diagnoses. Finally, this approach might be an effective way to inform a specialized systematic review where the search tools identify the cohort of potential studies to include which then undergo detailed review to extract important relationships between diagnoses.

**Conclusion**

Diagnostic errors can pose a serious threat to patients. This paper mines patterns from thousands of published misdiagnosis reports in the biomedical literature and structures the derived information in the form of a directed graph with frequency-weighted edges. It offers an additional way to understand misdiagnosis from a clinical or diagnostic decision support viewpoint The resulting materials are made available to the research community to inform clinical practice and research.